\documentclass{sf2a-conf2012}
\usepackage{hyperref}

\begin{document}

\TitreGlobal{SF2A 2012}

%%-----------------------------------------------------------------
%%      the top matter
%%

\title{Gaia and the dynamics of the Galaxy}

\runningtitle{Gaia and the dynamics of the Galaxy}

\author{B. Famaey}\address{Observatoire Astronomique, Universit\'e de Strasbourg, CNRS UMR 7550, F-67000 Strasbourg, France}

%% Keep this line, even if the page will be settled afterwards.
\setcounter{page}{1}

%%-----------------------------------------------------------------

\maketitle

%%-----------------------------------------------------------------
%%        The abstract
%% 
%%  Warning!  within the abstract:
%%  - do not use macros. 
%%  - do not use commands like: \cite, \citet, \citep ... etc.

\begin{abstract}
Gaia is an ambitious ESA space mission which will provide photometric and astrometric measurements with the accuracies needed to produce a kinematic census of almost one billion stars in our Galaxy. These data will revolutionize our understanding of the dynamics of the Milky Way, and our knowledge of its detailed gravitational potential and mass distribution, including the putative dark matter component and the non-axisymmetric features such as spiral arms. The Gaia mission will help to answer various currently unsettled questions by using kinematic information on both disk and halo stellar populations. Among many others: what does the rotation curve of the outer Galaxy look like? How far from axisymmetry and equilibrium is the Galaxy? What are the respective roles of hierarchical formation and secular evolution in shaping the Galaxy and its various components? Are the properties of the Galaxy in accordance with expectations from the standard model of cosmology?
\end{abstract}

%% Insert the keywords (to appear in the ADS indexing)
%% Keywords must be separated by a comma

\section{Introduction}
%---------------------

The Milky Way is a unique laboratory in which to test our models of galaxy formation, structure, and evolution. The  story of the efforts to obtain stellar kinematic data in the solar neighbourhood during the 20th century has culminated with ESA's Hipparcos astrometric catalogue (Perryman et al. 1997), and with the complementary ground-based spectroscopic surveys that have provided the missing information on the line-of-sight velocities (e.g., Nordstr\"om et al. 2004, Famaey et al. 2005). In recent years, drastic improvements in acquisition of stellar kinematic data outside of the close solar vicinity have been made through various large spectroscopic surveys, including the RAVE survey (Siebert et al. 2011a), as well as, for instance, SEGUE, SEGUE-2 and APOGEE, as part of SDSS (Eisenstein et al. 2011). Some of these newest data might actually slightly complicate our picture of the Galaxy, through possible signatures of significant deviations from axisymmetry and equilibrium. However, their interpretation is yet made difficult by sometimes complicated selection functions, and some of these data could also still be plagued by systematic errors, currently preventing us from drawing really secure conclusions. One of the most important sources of error could come from the distances of stars. In view of this, having an astrometric mission dedicated to parallax and proper motion measurements covering the whole sky will be of prime importance in order to refine our detailed understanding of the dynamics of the Galaxy. This is mainly what the Gaia mission will be about. \\

Scheduled to be launched from Kourou at the end of 2013, ESA's Gaia satellite (see, e.g., Prusti 2012) will perform astrometry and photometry of more than 1 billion objects up to a magnitude $V \sim 20$, as well as spectrosocopy for some 150 million objects up to $V \sim 16$. In order to get six-dimensional phase-space information for the whole Galaxy, the mission will have to be completed with ground-based spectrosocpic surveys, such as 4MOST and WEAVE, for stars  with $16<V<20$. Although there will be a plethora of science topics addressed by the mission, its primary objective will be to provide a detailed mapping of the Galaxy, and to develop, through dynamical means, a precise mass model including all of its components (interstellar gas, various stellar populations, as well as the yet hypothetical dark matter component needed in Newtonian gravity), giving us insight into its formation and evolutionary history. \\

This main objective will be achieved through dynamical models of the Galaxy. Such models have to rely on assumptions, such as axisymmetry and equilibrium. These very legitimate first order assumptions allow us to make use of Jeans' theorem constraining the phase-space distribution function to depend only on three isolating integrals of motion. There are various ways in which to compute these integrals (e.g., Binney 2012, Bienaym\'e \& Traven 2012). In this context, the action integrals are the best suited, since they are adiabatically invariant and are, with their conjugate angle variables, the natural coordinates of perturbation theory. By constructing distribution functions depending on these action integrals (e.g., Binney 2010), one can, e.g., disentangle the various stellar populations such as the thin and thick disk in a much more reliable way than with naive kinematic decomposition. In principle, one can iterate the fits with different gravitational potentials until the best-fitting potential is found, giving access to the underlying mass distribution, and hints into the missing mass question. It is also possible to take into account the main non-axisymmetric component (e.g., the bar) by modelling the system in its rotating frame. The other non-axisymmetric components should then be treated through perturbations. In the presence of multiple perturbers with significant amplitude, it is however not clear that this approach would work. In general, the consequences of assuming dynamical equilibrium when it is not the case might bias the results from the 0th order approach, assuming axisymmetry and equilibrium. In this respect, it will be extremely useful to test the Jeans approach on non-equilibrium models from simulations. The answer to such theoretical questions will have an important impact on the interpretation of future kinematical data. \\

Here, we list a series of questions that are still unsettled and sometimes overlooked when dealing with stellar kinematic data in our Galaxy. These questions will only be answered through a joint theoretical and observational effort, and using datasets with well-understood errors and selection functions, such as those that Gaia will provide. Answering these questions will certainly help refine our understanding of the structure, formation, and evolution of our Milky Way galaxy, and of galaxies in general.

\section{What does the rotation curve of the outer Galaxy look like?}

Rotation curves are the primary observational constraint on the gravitational field and the missing mass problem in external late-type galaxies (e.g., Gentile et al. 2004, Chemin et al. 2011). Paradoxically, even though one can obtain data with incomparable precision on the kinematics of individual stars in the Milky Way, our position inside the Galactic disk makes it difficult to measure the outer rotation curve with a similar precision as in external galaxies. This indeed requires to know our distance from the Galactic center $R_0$, the local circular velocity at the Sun's position $V_{c0}$ and the peculiar velocity of the Sun with respect to this circular velocity, as well as the precise distance to tracers used to measure the rotation curve in the outer Galaxy. Combining Gaia astrometric data with spectroscopic ones for the brightest stars, and with ground-based radial velocities for the faintest stars, should allow us to pin down the relevant quantities with the necessary precision. Note that the modelling should include an asymmetric drift correction for the tracer population. A possible caveat would however be that the non-axisymmetry of the Galaxy biases the estimation of the relevant parameters in an axisymmetric model (see Sect.~4). If the non-axisymmetry was too severe, the whole concept of a rotation curve would not even make sense anymore. \\

Nowadays, estimates of $R_0$ can vary, at the extremes, from $\sim$6.5~kpc to $\sim$9.5~kpc, while associated measurements of the local circular velocity vary from $\sim$180~km/s to $\sim$300~km/s, all these values being heavily model-dependent~(McMillan \& Binney 2010, Sch\"onrich 2012, Bovy et al. 2012a). Note that the currently often preferred value of $V_{c0}\sim 240$~km/s (see McMillan \& Binney 2010, Sch\"onrich 2012) would curiously make the Milky Way a significant fast-rotating outlier from the Tully-Fisher relation (see e.g. Flynn et al. 2006, Hammer et al. 2012 for a discussion). These values are however slightly degenerate with the peculiar motion of the Sun, and in particular with its azimuthal component, which is currently poorly constrained within a factor of five, lying in the interval between 5~km/s (Dehnen \& Binney 1998) and 25~km/s (Bovy et al. 2012a). \\

Independently of the absolute value of the circular velocity, another important question is the {\it shape} of the rotation curve in the outer Galaxy, which is crucial in determining the gravitational field and its associated putative dark matter distribution. It has long been argued that a dip at 9~kpc followed by a bump at 11~kpc is observed in the outer rotation curve of the Galaxy, which has notably been interpreted as the signature of the presence of ring-like dark matter structure (e.g., de Boer \& Weber 2011). Notwithstanding the fact that such a behavior could also be created by a dark matter halo with very large constant density core or by spiral arms, the interpretation of such a feature has also been attributed to a bias in the distribution of tracers (Binney \& Dehnen 1997), and the feature is not required in the recent dynamical modelling of APOGEE data (Bovy et al. 2012a). The case for the presence or absence of such a feature in the outer Galaxy rotation curve will certainly be settled with Gaia data, and will be of prime importance for ascertaining the presence or absence of dark matter substructures inside the Milky Way disk, in parallel with a better understanding of the precise density distribution of stars in the outer Galaxy, including features such as the Monoceros overdensity.

\section{Is the Galactic disk in vertical equilibrium?}

One of the great achievements of the Hipparcos mission in terms of Galactic dynamics (Cr\'ez\'e et al. 1998, Siebert et al. 2003, Holmberg \& Flynn 2004) has been to deduce the local dynamical mass in the Solar neighbourhood -- the Oort limit and the column density -- from the positions and velocities of tracer stars in the direction perpendicular to the Galactic plane, and relying on the Boltzmann and Poisson equations. Recently, using the kinematics of stars at heights of 1 to 4~kpc above the plane, the surface density as a function of height has been computed in the solar vicinity (Bovy \& Tremaine 2012), which allows to constrain the vertical structure of the dark matter halo. With parallax distances from Gaia, the precision on such measures will become watertight, and it should also be possible to directly measure the dynamical density in the disk as a function of position in the Galactic plane, allowing us to measure the dynamical scale-length and to compare it with the baryonic scale-length in order to test for the presence of a (real or effective) ``dark disk" component (Read et al. 2008, Bienaym\'e et al. 2009). \\

However, this relies on the {\it a priori} reasonable assumption that the disk is in vertical equilibrium. This assumption might be challenged by recent observations from SDSS (Widrow et al. 2012) measuring a 10\% North-South asymmetry in a 1~kpc-wide cylinder around the Sun comprising about 300000 stars: the asymmetry follows a wave-like behavior, with a South overdensity at heights of 500~pc and a North overdensity at 1~kpc. Complementary measurements of radial velocities for 11000 stars with SEGUE also reveal that the {\it mean} vertical motion can reach up to 10~km/s at heights of 1.5~kpc (to be compared with the assumed null mean vertical velocity in the plane yielding a solar reflex motion of $\sim$7~km/s), echoing previous similar results by Smith et al. (2012). If real, these nonzero mean vertical velocities of stars outside of the Galactic plane are very intriguing, and could be a signature of the Galactic warp, or of vertical perturbations excited by the recent passage of a satellite galaxy in the plane (Widrow et al. 2012). Alternatively, they could be due to self-excited vertical instabilities due to spiral arms, despite the presence of such bulk motions at relatively large heights above the plane.

\section{How far from axisymmetric is the Galaxy?}

Galactic disk instabilities, and their associated non-axisymmetric perturbations, including the bar and spiral arms, are known to be among the main drivers of the secular evolution of disks. Questions about their nature -- transient, quasi-stationary, or both types co-existing -- , about their detailed structure and dynamics -- e.g., amplitude and pattern speed -- , as well as questions about their influence on secular processes such as stellar migration (see Sect.~5), are all essential elements for a better understanding of Galactic evolution.

\subsection{Moving groups}

Six-dimensional phase-space information for stars in an increasingly large volume around the Sun will allow us to set new dynamical constraints on the non-axisymmetric perturbations of the Galactic potential. The resonances of the non-axisymmetric modes of the galactic potential are important agents locally disturbing the stellar velocity field. They create velocity substructures known as ``moving groups" in the Solar neighbourhood. These local velocity-space substructures have been reliably shown to be made of stars of very different ages and chemical compositions, so that the clumping cannot be due to irregular star formation (e.g., Dehnen 1998, Famaey et al. 2005, 2007, 2008, Pomp\'eia et al. 2011): they are nowadays the main stellar kinematical constraints on the non-axisymmetric components of the Galaxy. However, various models have argued to be able to represent these structures equally well, using transient or quasi-static spirals, with or without the help of the outer Lindblad resonance from the bar. However, all these models are making drastically different predictions on the velocity substructures when moving away from the Solar neighbourhood. Locating moving groups in velocity space outside of the solar neighbourhood is thus mandatory to discriminate between these models, and to obtain definitive constraints on the characteristics of the spiral arms and bar. The RAVE survey has for instance allowed (Antoja et al. 2012) to demonstrate that the Hercules moving group has a larger azimuthal velocity for regions inside the solar circle and a lower value outside, in accordance with what is expected from models where the Sun is located just outside the outer Lindblad resonance of the bar (e.g., Quillen et al. 2011).

\subsection{Oort constants}

The local effects of non-axisymmetric perturbations can also be analyzed by Taylor expanding to first order the planar velocity field in the cartesian frame of the Local Standard of Rest (LSR), an approximation roughly valid up to a distance of less than 2~kpc. This is done by generalizing the classical Oort constants to the case of a non-axisymmetric disk, yielding the constants $A$, measuring the azimuthal shear, $B$, measuring the vorticity, $C$, measuring the radial shear, and $K$, measuring the local divergence. Axisymmetry implies $C=K=0$ (but not the reverse, they can, e.g., be zero if the main non-axisymmetric perturbation is symmetric w.r.t. the Sun-Galactic center axis). Proper motions of a large sample of stars allow for a measurement of $A$, $B$, and $C$, while line-of-sight velocities projected onto the Galactic plane give access to $A$, $C$ and $K$. While old, rather imprecise, data were actually compatible with the axisymmetric values $C=K=0$ (Kuijken \& Tremaine 1994), a modern analysis of ACT/Tycho-2 proper motions, after corrections for the mode-mixing and asymmetric drift, yielded (Olling \& Dehnen 2003) $C = -10 \,$km/s/kpc for the red giants population, with a typical $\sigma_R \sim 40 \,$km/s. The measurement of the Oort constant $K$ was, on the other hand, recently performed with the RAVE survey (Siebert et al. 2011b) in the longitude interval $-140^\circ < l < 10^\circ$, thanks to the line-of-sight velocities (projected onto the Galactic plane) of 213713 stars (dominated by red giants) with spectro-photometric distances $d<2 \,$kpc from the Sun. This analysis confirmed the above proper-motion value of $C=-10$, and found a value $K= +6$, also different from zero. This value would actually imply a Galactocentric radial velocity gradient of $C+K = \partial V_R / \partial R \simeq - 4\,$km/s/kpc in the extended solar neighbourhood.

\subsection{Radial velocity gradient and LSR motion}

To check the actual existence of this radial velocity gradient implied by the above-measured value of $C+K$, the projection onto the plane of the mean line-of-sight velocity as  a function of $d \, {\rm cos} l \, {\rm cos} b$ for $|l|<5^\circ$, was examined both for the full RAVE sample and for red clump candidates (with an independent method of distance estimation).  The  observed mean velocities were then compared to  the expected velocities for  a thin disk  in circular rotation with an additional radial gradient. This clearly confirmed that the RAVE data  are  not compatible  with  a disk  in circular  rotation. They are, on the other hand, roughly compatible with a linear gradient of  $\partial V_R / \partial R \simeq - 4\,$km/s/kpc (Siebert et al. 2011b). This gradient in $\langle V_R \rangle$ is however not really linear. It is almost absent at small distances, and becomes steep at large distances from the Sun in the inner Galaxy. This means that it affects stars substantially above the plane, keeping in mind that RAVE lines of sight are typically at $b >\sim 20^\circ$: the zone where the gradient is steep concerns stars with $|z| \sim 500 \,$pc, located in the inner Galaxy and moving towards the anticenter. However, if the LSR itself is moving radially towards the inner Galaxy with $\sim 5$~km/s, then it would mean that the Galactic disk is locally affected by a motion towards the inner Galaxy, which slows down when one moves away from the Sun towards the Galactic center (see Fig.~4 in Famaey et al. 2012). Said in another way, one could have $\langle V_R \rangle \sim -5$~km/s in the Solar neighbourhood at $R \sim 8 \,$kpc, and $\langle V_R \rangle \sim 0$~km/s at $R \sim 6.5 \,$kpc along the GC-Sun axis and at $|z| \sim 500 \,$pc. This could explain the offset of 5~km/s between the local and non-local estimates of the radial motion of the Sun (Sch\"onrich 2012). \\

As a first step in understanding this gradient, one could assume that, to first order, what is seen slightly above the plane is a reflection of what would happen in a razor-thin disk, and that the spiral arms are tightly wound and long-lived (although this is heavily debated and most probably wrong to some extent), and described by the analytic Lin-Shu density wave model. This then allows us to constrain the shape, amplitude and dynamics of spiral arms, leaving the radial motion of the Sun as a free parameter. In Siebert et al. (2012) we showed that the best-fit model is obtained for a two-armed perturbation with the Sun close to the inner ultra-harmonic $4:1$ resonance, with a pattern speed $\Omega_s=18.6^{+0.3}_{-0.2} \,$km/s/kpc, and an amplitude of 14\% of the background density. In this fit the radial motion of the Sun stays within 1~km/s of its locally determined value. While very promising in giving a satisfactory fit, the model limitations nevertheless obviously prevent from drawing definitive conclusions about the cause of the gradient as well as the fitted parameters. It is for instance very likely that the bar also plays a role in this observed behavior.

\subsection{Implications}

Constraining the non-axisymmetries of the Galactic potential is very important for several reasons. First of all, interpreting the kinematic data in an axisymmetric model (see Sect.~2) could potentially lead to erroneous results, and thereby bias the estimate of the mass distribution in the outer Galaxy. Secondly, such non-axisymmetries are among the main drivers of the secular evolution of the Galactic disk (see Sect.~5), and can play an important role in driving stellar migrations, having dramatic consequences for the chemical evolution of the Galaxy. Thirdly, non-axisymmetric baryonic features, such as spiral arms, can be essential in breaking the disk-halo degeneracy: a large amplitude of the spiral structure cannot simply be traded for mass in the dark matter halo when fitting non-axisymmetric motions (see, e.g., Famaey \& Binney 2005 for a discussion). Finally, it is by disentangling the effects of the various non-axisymmetric components that the possible triaxiality of the dark matter halo could be spotted (or excluded) in the outer Galaxy. Such a triaxiality is predicted at large radii from the $\Lambda$CDM cosmological model. For this final point, the most promising approach is to use, as dynamical probes, the tidal streams of  disrupting satellite galaxies in the potential of the Milky Way, keeping in mind that these do not precisely delineate orbits. Kinematical information from Gaia should allow us to detect many such new streams in the stellar halo, and studying their shape and respective orbit is a very promising tool in constraining the shape of the Galactic potential at large radii. Many streams are indeed required for this method to be reliable, since the orbit of a single stream such as the Sagittarius stream (Ibata et al. 1994) could not break the degeneracy between the possible triaxiality of the halo and the shape of the rotation curve in the outer Galaxy. In this respect, combining dynamical constraints on streams with a reliable measure of the outer rotation curve (see Sect.~2) would be immensely useful for constraining the shape of the potential at large radii.

\section{What are the respective roles of hierarchical formation and secular evolution in shaping the Galaxy?}

One of the major goals of the dynamical modelling of stellar kinematical data obtained from Gaia will be to disentangle the various components of the Galaxy (bulge, stellar halo, thin disk, thick disk), and unravel their history and origin. Information on the chemical composition of the various stellar populations, obtained with ground-based spectroscopic surveys, will also be of prime importance in order to refine this understanding.

\subsection{The bulge and the stellar halo}

Two major processes are at play in the evolution of galaxies: hierarchical formation and secular evolution (see, e.g., Debattista et al. 2006 for an overview). The question of whether the Milky Way bulge is the result of an early major merger, of monolithic collapse, or simply of secular evolution through buckling of the central bar, or a combination of these processes, is still unsettled (Babusiaux et al. 2010). Similarly, we still do not know whether the stellar halo (see Helmi 2008 for a review) partially formed {\it in situ} through dissipational mergers of the first gaseous proto-galactic clumps, potentially resulting in a substantial fraction of its current inner regions, or  through the dissipationless accretion of a vast number of small galaxies (creating the aforementioned tidal streams in the outer halo, see Sect. 4). The existence of two such separate halo components and whether they could bear distinct kinematical signatures is still under heavy debate (Sch\"onrich et al. 2011, Beers et al. 2012). Obviously, the number of stellar streams newly detected with Gaia in the outer halo, as well as their phase-space characteristics and chemical composition (e.g., Prantzos 2011), will also allow us to quantify the role of hierarchical formation and merger rate in a cosmological context, testing the predictions of the $\Lambda$CDM model (see also Sect.~6).

\subsection{The thick disk}

On the other hand, the origin of the thick disk component is still a deep mystery: one possibility is that it was born {\it in situ}, e.g. from the internal gravitational instabilities in a gas-rich, turbulent, clumpy disk (Bournaud et al. 2009), or in the turbulent phase associated with numerous gas-rich mergers (Brook et al. 2004). It could also have been created through direct in-plane accretion of galaxy satellites, although modern data seem to rule out this extragalactic possibility (Ruchti et al. 2010, Wilson et al. 2011). The formation of the thick disk can either happen fast, on a Gyr timescale, in an early violent epoch, or as a secular process throughout the Galaxy lifetime. In the first case, the thick disk would appear as a clear distinct component in chemistry and phase-space, while in the latter case it would rather be seen as a gradual, continuous transition: modern data seem to favour this second option (Bovy et al. 2012b). On the other hand, studying the orbits of stars within the thick disk, notably their eccentricities at different galactocentric radii with Gaia data, will be a powerful tool for determining its origin (Sales et al. 2009, Di Matteo et al. 2011).

\subsection{Radial migration}

Driven by the absence of age-metallicity relation in the solar neighbourhood, chemical similarities between metal-poor bulge stars and the local thick disk (e.g., Mel\'endez et al. 2008), and the fact that stars labelled as local members of the thick disk can have orbital properties suggesting an origin at different galactocentric radii (Haywood 2008), there has recently been a growing convinction that stellar radial migration can result in the thick disk formation by bringing out stars with high velocity dispersion from the inner disk and the bulge (Sch\"onrich \& Binney 2009). Sellwood \& Binney (2002) indeed showed that transient spirals can efficiently redistribute angular momentum across their corotation radius without heating: if the amplitude of the perturber grows and decays on a timescale comparable to half the libration period of a horseshoe orbit around corotation, the spiral will deposit stars on the other side of corotation and vanish before pulling it back. In the presence of multiple patterns, horseshoe orbits are however necessarily destroyed at corotation when it overlaps with the Lindblad resonances of other patterns, unavoidably resulting in a very moderate degree of chaotic behaviour, even though this is disputed by some studies (Roskar et al. 2012). While differing from the previous process in contributing somewhat to the heating of the Galactic disk with time, this effect has been shown to also significantly enhance the efficiency of migration at corotation (Minchev \& Famaey 2010): in particular, in the presence of a strong central bar, transient spirals with corotation close to its outer Lindblad resonance are more efficient at mixing the disk, resulting in a peak of the redistribution of angular momentum at this radius (Minchev et al. 2011), revealing the important role of the bar for chemo-dynamical models taking into account stellar migration (see also Brunetti et al. 2011). For instance, the observed non-linearity of the metallicity gradient in the Milky Way, exhibiting at $R_0$ a step-like feature with overlapping metallicties characteristic of the inner and outer disk (L\'epine et al. 2011), might be directly linked to the coupling between the outer Lindblad resonance of the bar and the corotation of one of the Milky Way spiral patterns. Note also that the multiplicity of spiral patterns lead to the appearance of short-lived density peaks growing and decaying much faster than the spiral waves themselves, also greatly enhancing the efficiency of migration (Comparetta \& Quillen 2012). Generally speaking, stellar migration has, in any case, dramatic consequences on the chemo-dynamical evolution of the Milky Way disk, and of galaxy disks in general (e.g., Roskar et al. 2008). For instance, it flattens abundance gradients (probably more so in barred disks, Martin \& Roy 1994), and it creates extended stellar disks (up to 10 scale-lengths) with continuous (Type I) or downturning (Type II) density profiles (Minchev et al. 2012a), with steeper slopes for younger populations (Radburn-Smith et al. 2012). Additional smooth gas accretion could create additional spiral instabilities and more migration in the outer parts, thereby generating extended disks with upturning (Type III) profiles. In the Milky Way, it will be extremely interesting to check, with Gaia and complementary surveys such as WEAVE, whether an extended disk does exist, and whether it exhibits such signatures of radial migration, allowing us to quantify the roles of the non-axisymmetric components in the past chemo-dynamical evolution of the disk. \\

Sch\"onrich \& Binney (2009) have shown, with a simple chemical evolution model parametrizing radial migrations, that this mixing could explain most of the chemical properties of the thick disk. It was subsequently shown in Tree-SPH and cosmological simulations that stellar migrators tend to create a flared disk component rather than a thick disk of constant scale-height (Minchev et al. 2012b), meaning that radial migration does {\it not} thicken the disk in the inner parts. Given the hierarchical formation of the Galaxy, it is however also possible that the thick disk might have been formed without too much flaring through a combination of heating and radial migration caused by instabilities linked to mergers at high-redshift (Minchev et al. 2012c), i.e. quite rapidly in comparison with the Sch\"onrich \& Binney (2009) scenario but still in accordance with the Bovy et al. (2012b) results. Testing whether the thick disk has a constant scale-height as a function of radius, as in external galaxies, or whether it is rather strongly flared, would be very important to check whether secular evolution from radial migration alone can explain the creation of such a component. In that case, the scale-height should not be constant, and vertical chemical gradients should also vary with radius, while estimates of the average vertical action (see Sect.~1) for different populations of stars should reveal different variation with radius (flattening for older groups of stars). 

\section{Is the global structure of the Galaxy in accordance with cosmological expectations?}

\subsection{The Milky Way in a cosmological context}

Quantifying the hierarchical formation history of the Galaxy from understanding its various components as outlined above (notably the respective role of minor and major mergers in the formation of the bulge, thick disk and stellar halo) is of fundamental importance in order to test the predictions of the current standard $\Lambda$CDM cosmological model at low redshift. It should however be remembered that, on galaxy scales, predictions of this model are plagued by the enormous complications of baryonic astrophysics (e.g., Silk \& Mamon 2012). With this in mind, Milky Way observations nevertheless provide a unique, extremely useful tool in order to set constraints, and to present observational challenges that must be addressed by any scenario of galaxy formation. For instance, the existence of a large number of extended thin disk galaxies without classical bulges is in itself a challenge to the expectations from the number of major mergers in the cosmological model. Thus, tracing back the origin of the Milky Way bulge (Babusiaux et al. 2010 and Sect.~5 hereabove) would be extremely useful to set a benchmark allowing to compare it to the bulges of other galaxies, in order to determine their merger, bar-induced, or mixed origins. Similarly, a slow secular origin for the thick disk might be challenging to the number of expected mergers at high redshift in the standard cosmological scenario, and determining the exact origin of the thick disk and its relation with other components of the Galaxy is thus extremely important in this context (see Sect.~5).

\subsection{Dark matter distribution in $\Lambda$CDM}

A long-standing problem of $\Lambda$CDM is the fact that the numerical simulations of the collapse of dark matter halos lead to a density distribution as a function of radius, $\rho$, which is well fit by a smooth function asymptoting to a central cusp with slope ${\rm dln}\rho/{\rm dln}r \propto -r^{(1/n)}$ in the central parts (with $n \sim 6$ for a Milky Way-sized halo, meaning that the slope is still $-1$ at 200~pc from the center), while rotation curves of external galaxies lead to values of $n$ that are much closer to a constant density core (Chemin et al. 2011). In the Milky Way, non-axisymmetric motions (see Sect.~4) and microlensing events in the disk leave no room to trade mass from the stellar disk to the dark matter halo, and rule out a cuspy distribution (see, e.g., Famaey \& Binney 2005). Confirmation of this current state of affairs is expected from Gaia. The state-of-the-art solution to this problem is to enforce strong supernovae outflows that move large amounts of low-angular-momentum gas from the central parts and that pull on the central dark matter concentration to create a core (Governato et al. 2012), but this is still a relatively fine-tuned process, which fails to account for cored profiles in the smallest galaxies, and fails to produce their observed baryon fractions ([stars+gas]/total).

\subsection{Satellite galaxies: paucity and geometry}

Satellite galaxies of the Milky Way are also extremely useful tools to test the $\Lambda$CDM predictions. For instance, a lot of low-mass satellites are still missing (e.g., Bovill \& Ricotti 2011), which could be due to incomplete sky coverage from current surveys such as the SDSS. Discovering more dwarf galaxies around the Milky Way, especially ultra-faint ones is thus of immense importance for future surveys including Gaia. Sooner, the Pan-STARRS survey should already give us interesting results in this respect\footnote{Independently of future surveys, note that the most massive subhaloes currently predicted by $\Lambda$CDM are incompatible with hosting any of the MW satellites: the ``too big to fail" challenge (Bovill \& Ricotti 2011, Boylan-Kolchin et al. 2012)}. In addition to this challenge from the actual quantity of observed satellites, the distribution of dark subhalos around Milky Way-sized halos is also predicted by $\Lambda$CDM to be isotropic, or moderately flattened (Wang et al. 2012). However, the Milky Way satellites are currently observed to be highly correlated in phase-space: they lie within a seemingly rotationally supported thin disk (see, e.g., Kroupa et al. 2010). Young halo globular clusters define the same disk, and streams of stars and gas, roughly tracing the orbits of the objects from which they are stripped, seem to preferentially lie in this disk, too. While this might perhaps be explained by the infall of one or two groups of galaxies that would have retained correlated orbits, this solution is challenged by the fact that no nearby groups are observed to be anywhere near as spatially small as the Milky Way's disk of satellites (e.g., Kroupa et al. 2010). But since the SDSS survey covered only one fifth of the sky, it will of course be most interesting to see whether future surveys such as Pan-STARRS and Gaia will confirm this state of affairs. Precise proper motions observed with Gaia for the faintest galaxies and stellar streams will also be extremely important: this will allow us to trace back the orbits in a $\Lambda$CDM context (see, e.g., Angus et al. 2011), and will allow us to check whether the whole polar structure is indeed rotationally supported. Whether or not such a satellite phase-space correlation would be unique to the Milky Way should also be carefully checked: the Milky Way can be a statistical outlier, but if the $\Lambda$CDM model is a realistic description of nature, then the average satellite configurations in external galaxies should be only moderately flattened (Wang et al. 2012). If this is not the case, i.e. if flattened configurations are the norm, one possibility to investigate in detail would be a heavily heterogeneous and asymmetric reionization coming from internal sources, in the spirit of Ocvirk \& Aubert (2011), preserving only small regions of the proto-galaxy corresponding to the current disk of satellites.

\subsection{Modified gravity?}

Another more radical solution to this disk-of-satellites problem\footnote{Note that to maintain such a coherent rotating structure, the predicted triaxiality (see Sect.~4) of dark matter halos at large radii would also be severely challenged.} might be that most satellite galaxies are actually not primordial galaxies but rather tidal dwarf galaxies created in a major merger event, having occured in the Milky Way (e.g., Pawlowski et al. 2012) or even in M31 (Fouquet et al. 2012): this could possibly account for their presently correlated phase-space distribution, but would be in severe conflict with the fact that tidal dwarf galaxies are not supposed to harbour large dark matter fractions, while observations point to these objects being the most dark matter dominated objects in the Universe. This could however be reminiscent of young tidal dwarf galaxies observed in the tidal tails of external galaxies to also harbour substantial amounts of missing mass (Bournaud et al. 2007). One obvious solution might be to seriously reconsider the possibility that a large fraction of missing mass in disk galaxies is in the form of cold molecular gas distributed fractally in the disk (e.g., Pfenniger \& Combes 1994, Davies 2012). This would need us to rethink most of our understanding of galaxy formation. Another, even more baffling possibility, would be that there is actually no missing mass on galaxy scales (which would {\it not} mean that dark matter does not exist on cosmological scales), but that gravity is non-Newtonian. This would help explaining why many observed scaling relations in external galaxies (Famaey \& McGaugh 2012) involve the ubiquitous appearance of an acceleration constant $a_0 \sim \Lambda^{1/2} \sim 10^{-10}{\rm m} \, {\rm s}^{-2}$, whose origin is mysterious in the standard context. Most of these scaling relations can indeed be summarized by the empirical formula of Milgrom (1983), the success of which means that the gravitational field in galaxies mimicks, for whatever reason, an effectively modified force law on galaxy scales, known as Modified Newtonian Dynamics (MOND). This theory would for instance naturally explain why young tidal dwarf galaxies exhibit a mass discrepancy (Gentile et al. 2007) where dark matter is not expected to show up. The advent of Gaia will yield precise measurements allowing us to test the possibility that the success of Milgrom's formula is linked to modified gravity. These include the measurement of the outer rotation curve and its possible deviation from the baryonic Tully-Fisher relation (see Sect.~2), as well as the various effect of the ``phantom dark disk" of MOND (Bienaym\'e et al. 2009) on vertical velocity dispersions (see Sect.~3) and on the tilt of the stellar velocity ellipsoid, the precise shape of tidal streams around the Galaxy (see Sect.~4), or the effects of the external gravitational field in which the Milky Way is embedded on fundamental parameters such as the local escape speed. All these predictions can however slightly vary depending on the exact formulation of MOND, and will heavily depend on our precise knowledge of the underlying baryonic mass distribution. Should these tests be compatible with the predictions of MOND, the biggest challenge for such a theory would however remain to design a cosmological framework that could compete with the successes of $\Lambda$CDM on cosmological scales.

\section{Conclusion}

Today, there are still a lot of largely unknown answers to many old questions regarding the dynamics of the Milky Way, such as the local circular speed at the Sun's radius, the peculiar motion of the Sun, the degree of non-axisymmetry of the gravitational potential, the degree of North-South symmetry around the Galactic plane, or the geometry of the satellite system of the Galaxy. Recent data coming from large surveys are questioning many of the old assumptions that were made regarding these issues. Since it is still possible that some of these new results may turn out to be caused by systematic errors, notably in the stellar spectrophotometric distances, or from improper correction for complicated selection biases, it will be of prime importance to re-investigate all these questions with the distances and proper motions from Gaia at hand. Whatever the final answers to all these questions are, they will surely help us refine our understanding of the evolution of galaxies, and will give us new insights into modern questions about the gravitational field of the Milky Way and its connection to the dark matter mystery and cosmology. 

\begin{acknowledgements}
\end{acknowledgements}

%%-----------------------------
%%   Bibliography
%%-----------------------------

\end{document}